\documentclass[a4paper, 11pt]{amsart}
\usepackage{amsmath, amssymb, amsthm}	
\usepackage[foot]{amsaddr}				
\usepackage{braket, stmaryrd,mathrsfs}
\usepackage{cite}
\usepackage[all]{xy}
\usepackage{youngtab}					
\usepackage[draft]{graphicx}

\calclayout

\newcommand{\pr}[1]{#1^{\prime}}

\newcommand{\del}{\partial}

\newcommand{\mfrak}[1]{\mathfrak{#1}}
\newcommand{\mcal}[1]{\mathcal{#1}}
\newcommand{\mbb}[1]{\mathbb{#1}}
\newcommand{\mrm}[1]{\mathrm{#1}}

\newcommand{\scr}[1]{\mathscr{#1}}
\newcommand{\what}[1]{\widehat{#1}}

\theoremstyle{plain}
\newtheorem{thm}{Theorem}[section]

\newtheorem{cor}[thm]{Corollary}
\theoremstyle{definition}
\theoremstyle{remark}

\makeatletter
    
    \@addtoreset{equation}{section}
\makeatother

\title[Coset construction and WZW/SLE]{Coset construction of Virasoro minimal models and coupling of Wess-Zumino-Witten theory with Schramm-Loewner evolution}
\author{Shinji Koshida}
\address{Department of Basic Science, The University of Tokyo}
\email{koshida@vortex.c.u-tokyo.ac.jp}

\begin{document}

\begin{abstract}
Schramm-Loewner evolution (SLE) is a random process that gives a useful description of fractal curves.
After its introduction, many works concerning the connection between SLE and conformal field theory (CFT) have been carried out.
In this paper, we develop a new method of coupling SLE with a Wess-Zumino-Witten (WZW) model for $SU(2)$, an example of CFT,
relying on a coset construction of Virasoro minimal models.
Generalizations of SLE that correspond to WZW models were proposed by previous works
[Bettelheim {\it et al.}, Phys. Rev. Lett. {\bf 95}, 251601 (2005)] and [Alekseev {\it et al.}, Lett. Math. Phys. {\bf 97}, 243-261 (2011)],
in which the parameters in the generalized SLE for $SU(2)$ were related to the level of the corresponding $SU(2)$-WZW model.
The present work unveils the mechanism of how the parameters were chosen,
and gives a simpler proof of the result in these previous works,
shedding light on a new perspective of SLE/WZW coupling.
\end{abstract}

\maketitle

\section{Introduction}
After the introduction by Schramm \cite{Schramm2000}, an amount of studies on Schramm-Loewner evolution (SLE) has been carried out 
in probability theoretical manners \cite{Lawler2004,RohdeSchramm2005,LawlerSchrammWerner2001a,LawlerSchrammWerner2001b,LawlerSchrammWerner2002a,LawlerSchrammWerner2002b,Werner2003}.
SLE is a random family of uniformization maps of simply connected domains,
which is also equivalent to a random curve in a simply connected domain with end points anchored on the boundary.
In a typical case, this random curve is identified with an interface separating clusters in a critical lattice model such as the Ising model at the scaling limit \cite{Smirnov2001,ChelkakDuminil-CopinHonglerKemppainenSmirnov2014}.
Such a critical system is also known to be described by the framework of two dimensional (boundary) conformal field theory (CFT) \cite{BelavinPolyakovZamolodchikov1984,Cardy1986,Cardy1989}.
In CFT, it is often possible to exactly evaluate correlation functions due to its large symmetry described by a chiral algebra,
and this gives an exact prediction on a critical system.
A relevant example in our context is Cardy's formula \cite{Cardy1992, Smirnov2001}, which predicts the crossing probability in the two dimensional critical percolation.

Since SLE and CFT are two frameworks approaching critical systems, they could be connected to each other.
This direction of research is called SLE/CFT correspondence,
and the relation between SLE and the Virasoro algebra has been established by means of several formulations.
One of them is the group theoretical formulation originated by Bauer and Bernard \cite{BauerBernard2002, BauerBernard2003a,BauerBernard2003b},
which identified SLE with a random process on an infinite dimensional Lie group
and clarified the mechanism of computing local martingales associated with SLE relying on the representation theory of the Virasoro algebra.
We will shortly review this in Sect. \ref{sect:SLE_Virasoro}.
Other approaches include the one developped by Friedrich, Werner, Kontsevich, and Suhov \cite{FriedrichWerner2003, FriedrichKalkkinen2004,Friedrich2004,Kontsevich2003,KontsevichSuhov2007},
in which SLE was considered as a random process on a moduli space of Riemann surfaces via the Virasoro uniformization,
and a generalization of SLE to a random deformation of Riemann surfaces of higher genus was naturally formulated.
In more recent works by Dub\'{e}dat \cite{Dubedat2015a,Dubedat2015b}, the SLE partition function was constructed as a function on a Teichmuller space by a localization technique.

Along the line of SLE/CFT correspondence, several generalizations of SLE have been proposed.
For example, multiple SLE \cite{BauerBernardKytola2005} corresponds to multiple insertion of primary fields.
The most general framework of SLE connected to the Virasoro algebra was unified in \cite{Kytola2007}.
Other examples of generalization correspond to CFTs governed by other chiral algebras than the Virasoro algebra,
such as affine Lie (super) algebras \cite{BettelheimGruzbergLudwigWiegmann2005,AlekseevBytskoIzyurov2011,Sakai2013, SK2017,SK2018a,SK2018b} and superconformal algebras \cite{Rasmussen2004a, NagiRasmussen2005,SK2018c}.
A relevant example in the context of the current work is SLE with internal degrees of freedom
that corresponds to an $SU(2)$ Wess-Zumino-Witten (WZW) model governed by the affine $\mfrak{sl}_{2}$,
an origin of which we discover in the present paper.

To make the issue we address in the present paper concrete, we begin with the connection between usual SLE and the Virasoro algebra \cite{BauerBernard2002, BauerBernard2003a,BauerBernard2003b,Kytola2007}.
The SLE $\{g_{t}(z)\}_{t\ge 0}$ of parameter $\kappa$ (SLE($\kappa$) for short)
is a stochastic version of a Loewner evolution satisfying the stochastic Loewner equation
\begin{equation}
	\label{eq:stoch_Loewner}
	\frac{d}{dt}g_{t}(z)=\frac{2}{g_{t}(z)-\sqrt{\kappa}B_{t}}
\end{equation}
under the initial condition $g_{0}(z)=z$, where the solution is hydrodynamically normalized;
$g_{t}(z)=z+2t/z+o(z^{-1})$ as $z\to \infty$.
Here $B_{t}$ is the standard Brownian motion on the real axis starting at the origin.
In connection to CFT, it is also convenient to shift SLE by the Brownian motion: $f_{t}(z)=g_{t}(z)-\sqrt{\kappa}B_{t}$.
Then SLE is coupled to CFT via the following correlation function:
\begin{equation}
	\label{eq:martingale_Virasoro}
	\mcal{M}_{t}^{\mrm{Vir}}=\prod_{i}(\pr{f}_{t}(z_{i}))^{h_{i}}\braket{\Psi_{h_{\infty}}(\infty)\Psi_{h_{1}}(f_{t}(z_{1}))\cdots \Psi_{h_{n}}(f_{t}(z_{n}))\Psi_{h}(0)},
\end{equation}
where $\Psi_{\Delta}$ is the primary field of conformal weight $\Delta$
and the expectation value is taken for CFT of central charge $c$.
The SLE/CFT correspondence \cite{BauerBernard2002, BauerBernard2003a,BauerBernard2003b,Kytola2007}
states that the above random process $\mcal{M}_{t}^{\mrm{Vir}}$ is a local martingale
if the central charge and the conformal weight inserted at the origin are chosen as
\begin{equation}
	\label{eq:coupling_Virasoro}
	c=c_{\kappa}=1-\frac{3(\kappa-4)^{2}}{2\kappa},\ \ \ h=h_{\kappa}=\frac{6-\kappa}{2\kappa}.
\end{equation}

A generalization of SLE that is coupled to a WZW model for a compact simple Lie group $G$
was proposed in \cite{BettelheimGruzbergLudwigWiegmann2005,AlekseevBytskoIzyurov2011},
and studied by the present author \cite{SK2017,SK2018a} relying on an infinite dimensional Lie group.
Let us denote the complexification of the Lie algebra of $G$ by $\mfrak{g}$
and denote a finite dimensional irreducible representation of $\mfrak{g}$ of highest weight $\Lambda$ by $L(\Lambda)$.
The approach by \cite{BettelheimGruzbergLudwigWiegmann2005,AlekseevBytskoIzyurov2011} began from considering a random process
\begin{equation}
	\label{eq:martingale_WZW}
	\mcal{M}_{t}^{G}=\prod_{i}(\pr{f}_{t}(z))^{h_{\lambda_{i}}}\braket{\Psi(\ket{\lambda_{\infty}},\infty)\Psi(\ket{\lambda_{1}(z_{1})}_{t},f_{t}(z_{1}))\cdots \Psi(\ket{\lambda_{n}(z_{n})}_{t},f_{t}(z_{n}))\Psi(\ket{v_{\lambda}},0)},
\end{equation}
where $\Psi(\ket{v_{\Lambda}},z)$ is the primary field specified by a vector $\ket{v_{\Lambda}}$
lying in $L(\Lambda)$ for a weight $\Lambda$ of $\mfrak{g}$, and its conformal weight is denoted by $h_{\Lambda}$.
The expectation value is taken for the WZW model for $G$ with level $k$.
Again the random process $f_{t}(z)$ is the SLE$(\kappa)$ with a shift by the Brownian motion and
a random process $\ket{\lambda(z)}_{t}$ on $L(\lambda)[[z^{-1}]]$ satisfies the following stochastic differential equation (SDE):
\begin{equation}
	\label{eq:SDE_internal}
	d\ket{\lambda(z)}_{t}=\frac{\tau}{2}\sum_{a=1}^{\dim\mfrak{g}}\left(\frac{X_{a}}{f_{t}(z)}\right)^{2}\ket{\lambda(z)}_{t}dt-\sqrt{\tau}\sum_{a=1}^{\dim\mfrak{g}} \frac{X_{a}}{f_{t}(z)}\ket{\lambda(z)}_{t}dW_{t}^{(a)}.
\end{equation}
Here $W_{t}^{(a)}$ are mutually independent standard Brownian motions that are also independent of
the Brownian motion $B_{t}$ appearing in the stochastic Loewner equation in Eq.(\ref {eq:stoch_Loewner}),
and $\{X_{a}\}_{a=1}^{\dim\mfrak{g}}$ is an orthonormal basis of $\mfrak{g}$ with respect to the normalized nondegenerate symmetric invariant bilinear form on $\mfrak{g}$.
It was argued in \cite{BettelheimGruzbergLudwigWiegmann2005,AlekseevBytskoIzyurov2011} that
if $\mfrak{g}=\mfrak{sl}_{2}$, $\lambda$ is the fundamental weight, and the parameters $\kappa$ and $\tau$ are chosen as
\begin{equation}
	\kappa=4\frac{k+2}{k+3},\ \ \  \tau=\frac{2}{k+3},
\end{equation}
then the random process $\mcal{M}_{t}^{G}$ was a local martingale.

In \cite{BettelheimGruzbergLudwigWiegmann2005}, it was pointed out that for this $\kappa$,
the corresponding central charge given in Eq.(\ref{eq:coupling_Virasoro}) for the Virasoro algebra is
\begin{equation}
	c_{\kappa}=1-\frac{6}{(k+2)(k+3)},	
\end{equation}
which is identical to the central charge for the Virasoro algebra
obtained by a coset construction denoted as $(SU(2)_{k}\times SU(2)_{1})/SU(2)_{k+1}$.

The initial motivation of the current work was to better understand this phenomenon;
the appearance of central charge for a coset construction,
but studies in this direction revealed an origin of SLE with internal degrees of freedom
that is coupled to WZW model for $SU(2)$ in an unexpected way,
and how the parameter $\tau$ is chosen.
We also provide a simpler proof of the assertion that
$\mcal{M}_{t}^{G}$ is a local martingale for a certain choice of $\kappa$ and $\tau$.

The organization of the present paper is as follows:
The next Sect. \ref{sect:vertex_algebra} gives preliminaries on the relevant vertex algebras in the current work.
It contains exposition on Virasoro vertex algebras and affine vertex algebras for $\mfrak{sl}_{2}$.
We also recall the coset construction of Virasoro vertex algebras developed in \cite{GoddardKentOlive1986, KacWakimoto1988}.
In Sect. \ref{sect:SLE_Virasoro}, we review the group theoretical formulation of SLE originated by Bauer and Bernard \cite{BauerBernard2002, BauerBernard2003a,BauerBernard2003b},
which connects SLE to the representation theory of the Virasoro algebra.
In Sect. \ref{sect:SLE_WZW}, we discuss a natural extension of SLE that is coupled to WZW theory for $SU(2)$
from the viewpoint of the coset construction.
In the final Sect. \ref{sect:conclusion}, we conclude the present paper and make some discussion.

\section*{Acknowledgements}
The author thanks R. Sato for information on the coset construction of Virasoro minimal models.
He is also grateful to M. Katori for discussions on related topics to the content of the present paper.
This work was supported by a Grant-in-Aid for JSPS Fellows (Grant No. 17J09658).

\section{Preliminaries on vertex algebras}
\label{sect:vertex_algebra}
In this section, we give a brief introduction to vertex algebras
that are used in the subsequent sections,
and review the coset construction \cite{GoddardKentOlive1986, KacWakimoto1988} of Virasoro vertex algebras.
A more detailed exposition can be found in books \cite{Kac1998, FrenkelBen-Zvi2004,IoharaKoga2011}.

\subsection{Virasoro vertex algebra}
One of the most fundamental examples of vertex algebras is the Virasoro vertex algebra.
Virasoro algebra is an infinite dimensional Lie algebra $\mrm{Vir}=\bigoplus_{n\in\mbb{Z}}\mbb{C}L_{n}\oplus \mbb{C}C$
with Lie bracket given by
\begin{equation}
	[L_{m},L_{n}]=(m-n)L_{m+n}+\frac{m^{3}-m}{12}\delta_{m+n,0}C,\ \ \ [C,\mrm{Vir}]=\{0\}.
\end{equation}
To define relevant representations of the Virasoro algebra,
we take subalgebras $\mrm{Vir}_{\pm}=\bigoplus_{\pm n>0}\mbb{C}L_{n}$ and $\mrm{Vir}_{0}=\mbb{C}L_{0}\oplus \mbb{C}C$.
We also write $\mrm{Vir}_{\ge 0}$ for $\mrm{Vir}_{+}\oplus \mrm{Vir}_{0}$.
For a given pair $(c,h)\in\mbb{C}^{2}$, we define a one dimensional representation $\mbb{C}\ket{c,h}$ of $\mrm{Vir}_{\ge 0}$
by $L_{n}\ket{c,h}=0$ for $n>0$, $L_{0}\ket{c,h}=h\ket{c,h}$, and $C\ket{c,h}=c\ket{c,h}$.
Then the corresponding Verma module $M(c,h)$ is defined by
\begin{equation}
	M(c,h)=U(\mrm{Vir})\otimes_{U(\mrm{Vir}_{\ge 0})}\mbb{C}\ket{c,h}\simeq U(\mrm{Vir}_{-})\ket{c,h},
\end{equation}
where the latter isomorphism is one as vector spaces.
A Verma module is irreducible for a generic pair $(c,h)$,
but becomes reducible for a specific choice.
We denote the irreducible quotient by $L(c,h)$.
Here two parameters $c$ and $h$ are often called the central charge and the conformal weight of the corresponding representation, respectively.

On a Verma module $M(c,h)$, one can uniquely define a bilinear form $\braket{\cdot|\cdot}:M(c,h)\times M(c,h)\to \mbb{C}$
by imposing
\begin{equation}
	\braket{c,h|c,h}=1,\ \ \ \braket{L_{n}u|v}=\braket{u|L_{-n}|v},\ \ u,v\in M(c,h).
\end{equation}
This bilinear form descends to a nondegenerate one on the irreducible quotient $L(c,h)$.

Among others, an irreducible representation $L(c,0)$ has a special feature that
it admits a structure of a vertex algebra.
We write $L^{\mrm{Vir}}_{c}$ for this vertex algebra and call it the Virasoro vertex algebra of central charge $c$.
This vertex algebra is generated by a single conformal vector $\omega=L_{-2}\ket{0}$,
where $\ket{0}=\ket{c,0}$ is the vacuum vector, with the corresponding field
\begin{equation}
	Y(\omega,z)=L(z):=\sum_{n\in\mbb{Z}}L_{n}z^{-n-2}.
\end{equation}

A particular interest in the present paper is in the case when the central charge lies in the minimal series \cite{BelavinPolyakovZamolodchikov1984}
\begin{equation}
	c=c^{\mrm{min}}_{p,q}=1-\frac{6(p-q)^{2}}{pq}
\end{equation}
for comprime integers $p$ and $q$ larger or equal than $3$.
For this central charge, modules over the corresponding Virasoro vertex algebra
$L^{\mrm{Vir}}_{c^{\mrm{min}}_{p,q}}$ are exhausted by $L(c^{\mrm{min}}_{p,q},h_{p,q;r,s})$,
where
\begin{equation}
	h_{p,q;r,s}=\frac{(rq-sp)^{2}-(p-q)^{2}}{4pq}
\end{equation}
for $r=1,\cdots ,p-1$ and $s=1,\cdots q-1$.
Note that this list over-counts modules because the transformation $r\to p-r$, $s\to q-s$ preserves the corresponding conformal weight.

It is known that the quotient map $M(c^{\mrm{min}}_{p,q},h_{p,q;r,s})\to L(c^{\mrm{min}}_{p,q},h_{p,q;r,s})$
has a nontrivial kernel.
Indeed the Verma module $M(c^{\mrm{min}}_{p,q},h_{p,q;r,s})$ carries a singular vector
$\ket{\chi_{p,q;r,s}}$ that is an eigenvector of $L_{0}$ corresponding to
an eigenvalue $h_{p,q;r,s}+rs$.
A relevant example in our context is the case when $(r,s)=(2,1)$.
In this case the vector
\begin{equation}
	\ket{\chi_{p,q;2,1}}=\left(-2L_{-2}+\frac{\kappa}{2}L_{-1}^{2}\right)\ket{c^{\mrm{min}}_{p,q},h_{p,q;2,1}},
\end{equation}
where $\kappa=4p/q$ is a singular vector.

\subsection{Affine vertex algebra for $\mfrak{sl}_{2}$}
Though an affine vertex algebra is defined associated with an arbitrary finite dimensional reductive Lie algebra $\mfrak{g}$,
we concentrate on the case when $\mfrak{g}=\mfrak{sl}_{2}$.
We take a standard basis of $\mfrak{g}=\mfrak{sl}_{2}$
\begin{equation}
	E=\left(
		\begin{array}{cc}
			0	& 1\\
			0	& 0
		\end{array}	
		\right),\ \ \ 
	H=\left(
		\begin{array}{cc}
			1	& 0\\
			0	& -1
		\end{array}	
		\right),\ \ \ 
	F=\left(
		\begin{array}{cc}
			0	& 0\\
			1	& 0
		\end{array}	
		\right),
\end{equation}
and let $(\cdot|\cdot)$ be a nondegenerate symmetric invariant bilinear form on $\mfrak{g}$
defined by $(X|Y)=\mrm{Tr}(XY)$,
where the product and the trace are computed by regarding the entries as two by two matrices.

Then the corresponding affine Lie algebra is defined by
$\what{\mfrak{g}}=\mfrak{g}\otimes\mbb{C}[\zeta,\zeta^{-1}]\oplus \mbb{C}K$ with Lie bracket
\begin{equation}
	[X(m),Y(n)]=[X,Y](m+n)+m(X|Y)\delta_{m+n,0}K,\ \ \ [K,\what{\mfrak{g}}]=\{0\},
\end{equation}
where $X(m)=X\otimes \zeta^{m}$ for $X\in\mfrak{g}$.

Recall that finite dimensional irreducible representations of $\mfrak{g}$
are parametrized by non-negative half integers;
namely, for $j\in\frac{1}{2}\mbb{Z}_{\ge 0}$, the corresponding irreducible representation
$L(j)$ is $(2j+1)$-dimensional and generated by a highest weight vector $\ket{j}$ such that
$E\ket{j}=0$ and $H\ket{j}=2j\ket{j}$.
To consider representations of $\what{\mfrak{g}}$,
we extend the action of $\mfrak{g}$ on $L(j)$ to one of $\mfrak{g}\otimes \mbb{C}[\zeta]\oplus\mbb{C}K$ by
identifying $\mfrak{g}$ with the zero-mode subalgebra $\mfrak{g}\otimes \zeta^{0}$
and by the defining properties $(\mfrak{g}\otimes\mbb{C}[\zeta]\zeta) L(j)=\{0\}$
and $K\mapsto k\mrm{Id}$ for some complex number $k\in\mbb{C}$.
Then a representation of $\what{\mfrak{g}}$ is constructed as
\begin{equation}
	\what{L(j)}_{k}=U(\what{\mfrak{g}})\otimes_{U(\mfrak{g}\otimes\mbb{C}[\zeta]\oplus\mbb{C}K)}L(j)\simeq U(\mfrak{g}\otimes\mbb{C}[\zeta^{-1}]\zeta^{-1})\otimes L(j).
\end{equation}
This representation is not necessarily irreducible as a representation of $\what{\mfrak{g}}$,
and its irreducible quotient is denoted by $L_{\mfrak{g},k}(j)$.

On a representation $L_{\mfrak{g},k}(j)$, we define a nondegenerate bilinear form $\braket{\cdot|\cdot}$ by
\begin{equation}
	\braket{j|j}=1,\ \ \ \braket{X(n)u|v}=-\braket{u|X(-n)v},\ \ \ u,v\in L_{\mfrak{g},k}(j).
\end{equation}

We can equip a vertex algebra structure on a vacuum representation $L_{\mfrak{g},k}=L_{\mfrak{g},k}(0)$.
This vertex algebra is called the affine vertex algebra for $\mfrak{sl}_{2}$ of level $k$.
A particular interest is in the case when $k$ is admissible in the sense of Kac and Wakimoto \cite{KacWakimoto1989},
{\it i.e.},
\begin{equation}
	k=-2+\frac{p}{q}
\end{equation}
for coprime positive integers $p$ and $q$ such that $p\ge 2$.
Modules over $L_{\mfrak{g},k}$ for an admissible level $k$ were classified in \cite{KacWakimoto1989},
which we do not refer to in the present paper.
In the case when $k$ is a positive integer, modules over $L_{\mfrak{g},k}$ are exhausted by $L_{\mfrak{g},k}(j)$
for $0\le j\le k/2$ \cite{FrenkelZhu1992}.

In an affine vertex algebra, we have a conformal vector via the Sugawara construction.
Let $\{X_{a}\}_{a=1}^{\dim\mfrak{g}}$ be an orthonormal basis of $\mfrak{g}$ with respect to
the given bilinear form $(\cdot|\cdot)$.
For example, we can take
\begin{equation}
	X_{1}=\frac{1}{\sqrt{2}}H,\ \ X_{2}=\frac{1}{2}(E+F),\ \ X_{3}=\frac{i}{\sqrt{2}}(E-F).
\end{equation}
Then a vector defined by
\begin{equation}
	\omega^{\mrm{Sug}}_{k}:=\frac{1}{2(k+2)}\sum_{a=1}^{\dim\mfrak{g}}X_{a}(-1)^{2}\ket{0}
\end{equation}
is a conformal vector of central charge $c^{\mrm{Sug}}_{k}=\frac{3k}{k+2}$,
{\it i.e.}, the coefficients in the corresponding field
\begin{equation}
	Y(\omega^{\mrm{Sug}}_{k},z)=\sum_{n\in\mbb{Z}}L_{n}z^{-n-2}
\end{equation}
define a representation of the Virasoro algebra on $L_{\mfrak{g},k}$
and its modules $L_{\mfrak{g},k}(j)$.
The conformal weight of a representation $L_{\mfrak{g},k}(j)$ with respect to this Virasoro action
is given by $h^{\mrm{Sug}}_{j}=\frac{j(j+1)}{k+2}$.

\subsection{Coset construction of Virasoro minimal models}
Here we recall the notion of a commutant vertex algebra.
Let $V$ be a vertex algebra with a conformal vector $\omega_{V}$
and let $W\subset V$ be a vertex subalgebra with a conformal vector $\omega_{W}$.
For a vector $A\in V$, we expand the corresponding field so that
\begin{equation}
	Y(A,z)=\sum_{n\in\mbb{Z}}A_{(n)}z^{-n-1},
\end{equation}
where $A_{(n)}\in\mrm{End}(V)$.
Then the commutant vertex algebra denoted by $\mrm{Com}(W,V)$ is defined by
\begin{equation}
	\mrm{Com}(W,V)=\{v\in V|w_{(n)}v=0,\ w\in W,\ n\ge 0\}.
\end{equation}
This is a vertex algebra with a conformal vector $\omega_{V}-\omega_{W}$.

A particularly important example is the one developed in \cite{GoddardKentOlive1986,KacWakimoto1988},
which we explain now.
Firstly, we have an injective vertex algebra homomorphism $\iota:L_{\mfrak{g},k+1}\to L_{\mfrak{g},k}\otimes L_{\mfrak{g},1}$
defined by $X(-1)\ket{0}\mapsto (X(-1)\ket{0})\otimes\ket{0}+\ket{0}\otimes (X(-1)\ket{0})$.
Note that the image of the conformal vector $\omega^{\mrm{Sug}}_{k+1}$ in $L_{\mfrak{g},k+1}$
under the homomorphism $\iota$ is computed as
\begin{align*}
	2(k+3)\iota(\omega^{\mrm{Sug}}_{k+1})
	=&2(k+2)\omega^{\mrm{Sug}}_{k+1}\otimes \ket{0}+6\ket{0}\otimes \omega^{\mrm{Sug}}_{1}\\
	&+2\biggl(\frac{1}{2}H(-1)\ket{0}\otimes H(-1)\ket{0}+E(-1)\ket{0}\otimes F(-1)\ket{0}\\
	&\hspace{25pt}+F(-1)\ket{0}\otimes E(-1)\ket{0}\biggr).
\end{align*}
From a general principle, the corresponding commutant vertex algebra
$\mrm{Com}(L_{\mfrak{g},k+1},L_{\mfrak{g},k}\otimes L_{\mfrak{g},1})$ carries a conformal vector
$\omega^{\mrm{Com}}=\omega^{\mrm{Sug}}_{k}\otimes \ket{0}+\ket{0}\otimes \omega^{\mrm{Sug}}_{1}-\iota (\omega^{\mrm{Sug}}_{k+1})=-\frac{1}{k+3}\Omega$,
where $\Omega$ is the generalized Casimir vector given by
\begin{align*}
	\Omega=
	&\frac{1}{2}H(-1)\ket{0}\otimes H(-1)\ket{0}+E(-1)\ket{0}\otimes F(-1)\ket{0}+F(-1)\ket{0}\otimes E(-1)\ket{0} \\
	&-\omega^{\mrm{Sug}}_{k}\otimes \ket{0}-k\ket{0}\otimes\omega^{\mrm{Sug}}_{1}.
\end{align*}
Then the central charge of this conformal vector is
\begin{equation}
	c^{\mrm{Com}}_{k}=c^{\mrm{Sug}}_{k}+c^{\mrm{Sug}}_{1}-c^{\mrm{Sug}}_{k+1}=1-\frac{6}{(k+2)(k+3)}.
\end{equation}
Notice that in the case when $k=-2+p/q$ is admissible, this central charge coincides with $c^{\mrm{min}}_{p,p+q}$.
In fact, more strictly, an isomorphism $\mrm{Com}(L_{\mfrak{g},k+1},L_{\mfrak{g},k}\otimes L_{\mfrak{g},1})\simeq L^{\mrm{Vir}}_{c^{\mrm{min}}_{p,p+q}}$ holds.
Now let us recall the precise statement in \cite{KacWakimoto1988},
which in a particular case when $k$ is a positive integer reduces to the result in \cite{GoddardKentOlive1986}:
\begin{thm}[\cite{KacWakimoto1988}]
	\label{thm:GKO}
	Let $k=-2+p/q$ be an admissible level.
	A representation $L_{\mfrak{g},k}(j)\otimes L_{\mfrak{g},1}(\epsilon)$
	of $\mfrak{g}\oplus\mfrak{g}$ for $0\le j\le (p-2)/2$ and $\epsilon=0$ or $1/2$
	is decomposed as a representation of $\mrm{Vir}\oplus \mfrak{g}$ as follows:
	\begin{equation}
	L_{\mfrak{g},k}(j)\otimes L_{\mfrak{g},1}(\epsilon)=\bigoplus_{\substack{s=1\\ r-s\equiv 2\epsilon\ \mrm{mod}\ 2}}^{p+q-1}L(c^{\mrm{min}}_{p,p+q},h_{p,p+q;r,s})\otimes L_{\mfrak{g},k+1}\left(\frac{s-1}{2}\right),
	\end{equation}
	where $r=2j+1$.
\end{thm}

\section{Connection of SLE and the Virasoro algebra}
\label{sect:SLE_Virasoro}
This section is devoted to an exposition of the group theoretical formulation of SLE originated by Bauer and Bernard \cite{BauerBernard2002, BauerBernard2003a,BauerBernard2003b},
which explained how local martingales associated with SLE were computed from a representation of the Virasoro algebra.
\subsection{Infinite dimensional group $\mrm{Aut}_{+}\mcal{O}$}
Let $\mcal{O}=\mbb{C}[[z^{-1}]]$ be a completed topological $\mbb{C}$-algebra,
and let $\mrm{Aut}\mcal{O}$ be the group of continuous automorphisms of $\mcal{O}$.
Since each element $\rho\in\mrm{Aut}\mcal{O}$ is identified with an infinite series $\rho(z)$,
the group $\mrm{Aut}\mcal{O}$ is realized as
\begin{equation}
	\mrm{Aut}\mcal{O}\simeq \left\{a_{1}z+a_{0}+a_{-1}z^{-1}+\cdots \in \mbb{C}[[z^{-1}]]z |a_{1}\neq 0\right\}.
\end{equation}
We define a group law on this group by $(\mu\ast\rho)(z):=\rho(\mu(z))$ for $\rho,\mu\in\mrm{Aut}\mcal{O}$.
A particularly important subgroup in the present paper denoted by $\mrm{Aut}_{+}\mcal{O}$ is defined under this realization by
\begin{equation}
	\mrm{Aut}_{+}\mcal{O}\simeq \{z+a_{0}+a_{-1}z^{-1}+\cdots\}\simeq z+\mbb{C}[[z^{-1}]].
\end{equation}

We can consider the Lie algebras of these groups;
they consist of vector fields holomorphic at $z=\infty$ so that
$\mrm{Lie}(\mrm{Aut}\mcal{O})=\mrm{Der}_{0}\mcal{O}\simeq z\mbb{C}[[z^{-1}]]\del_{z}$,
and $\mrm{Lie}(\mrm{Aut}_{+}\mcal{O})=\mrm{Der}_{+}\mcal{O}\simeq \mbb{C}[[z^{-1}]]\del_{z}$.

In the context of SLE, the smaller Lie algebra $\mrm{Der}_{+}\mcal{O}$ and the corresponding group $\mrm{Aut}_{+}\mcal{O}$
are sufficient in formulation.
What will become important is that they act on a certain completion of a representation space of the Virasoro algebra.
Since the operator $L_{0}$ is diagonalized on a representation $L(c,h)$ so that
$L(c,h)=\bigoplus_{n=0}^{\infty}L(c,h)_{h+n}$,
where each eigenspace $L(c,h)_{h+n}$ corresponding to an eigenvalue $h+n$ is finite-dimensional.
Thus we can take the formal completion as the direct product;
$\overline{L(c,h)}:=\prod_{n=0}^{\infty} L(c,h)_{h+n}$.
For each element $\rho\in\mrm{Aut}_{+}\mcal{O}$, we uniquely find numbers $v_{i}$ for $i\le -1$ so that
\begin{equation}
	\exp\left(\sum_{j\le -1}v_{j}z^{j+1}\del_{z}\right)z=\rho(z).
\end{equation}
Then from these data, we define an operator $Q(\rho)$ by
\begin{equation}
	Q(\rho):=	\exp\left(-\sum_{j\le -1}v_{j}L_{j}\right),
\end{equation}
and find that this assignment $Q:\mrm{Aut}_{+}\mcal{O}\to \mrm{End}(\overline{L(c,h)})$ is a representation;
it satisfies $Q(\rho\ast\mu)=Q(\rho)Q(\mu)$.
Note that the completion of a representation space is required to make an operator $Q(\rho)$ well-defined.

\subsection{Null vector and local martingales}
A remarkable discovery by Bauer and Bernard \cite{BauerBernard2002, BauerBernard2003a,BauerBernard2003b} was that the shifted SLE
$f_{t}(z)=g_{t}(z)-\sqrt{\kappa}B_{t}$
can be identified with a random process on the infinite dimensional Lie group $\mrm{Aut}_{+}\mcal{O}$.
To recall this identification, we consider a random process $f_{t}$ on $\mrm{Aut}_{+}\mcal{O}$ that satisfies
the following SDE:
\begin{equation}
	f_{t}^{-1}df_{t}=\left(-2\ell_{-2}+\frac{\kappa}{2}\ell_{-1}^{2}\right)dt+\sqrt{\kappa}\ell_{-1}dB_{t},
\end{equation}
with the initial condition $f_{0}=\mrm{Id}$, where we write $\ell_{n}=-z^{n+1}\del_{z}$ and
$B_{t}$ is the standard Brownian motion on the real axis that starts at the origin.
Then we can find that the corresponding random process $f_{t}(z)$
with infinite series as its values satisfies the following SDE:
\begin{equation}
	df_{t}(z)=\frac{2}{f_{t}(z)}dt + \sqrt{\kappa}dB_{t},
\end{equation}
and $f_{0}(z)=z$.
Since the infinite series $f_{t}(z)$ is properly normalized,
we can identify the random process $\{f_{t}(z)\}_{t\ge 0}$ with the shifted SLE($\kappa$).

A virtue of this formulation of SLE as a random process on the infinite dimensional Lie group $\mrm{Aut}_{+}\mcal{O}$
is that it allows one to compute local martingales associated with SLE
from the representation theory of the Virasoro algebra.
Indeed, when we start from a random process $f_{t}$ on $\mrm{Aut}_{+}\mcal{O}$,
we can consider an operator valued random process $Q(f_{t})$.
Then a random process $Q(f_{t})\ket{c_{\kappa},h_{\kappa}}$ on $\overline{L(c_{\kappa},h_{\kappa})}$
is a local martingale due to the fact that the vector $(-2L_{-2}+\frac{\kappa}{2}L_{-1}^{2})\ket{c_{\kappa},h_{\kappa}}$
vanishes because it comes from a singular vector in the corresponding Verma module.

We can show that the local martingale presented in Eq.(\ref{eq:martingale_Virasoro})
is deduced from this vector-valued local martingale in the following three steps:
firstly, we regard the bracket $\braket{\ }$ as the vacuum expectation $\braket{0|\ |0}$.
Secondly, we use a property of a primary field applied to the vacuum vector that
$\ket{c_{\kappa},h_{\kappa}}=\Psi_{h_{\kappa}}(0)\ket{0}$.
Finally, we notice that a primary field is transformed under the adjoint action by $Q(\rho)$
for $\rho\in\mrm{Aut}_{+}\mcal{O}$ so that
\begin{equation}
	\Psi_{h}(z)=Q(\rho)\Psi_{h}(\rho(z))Q(\rho)^{-1}(\pr{\rho}(z))^{h}.
\end{equation}
Then we can identify the random process $\mcal{M}_{t}^{\mrm{Vir}}$ with
\begin{equation}
	\mcal{M}_{t}^{\mrm{Vir}}=\braket{0|\Psi_{h_{\infty}}(\infty)\Psi_{h_{1}}(z_{1})\cdots \Psi_{h_{n}}(z_{n})Q(f_{t})|c_{\kappa},h_{\kappa}}	,
\end{equation}
which is obviously a local martingale.

Note that in the case when $\kappa=4p/q$ for some coprime integers $p,q\ge 3$,
the corresponding central charge and conformal weight are given by
$c_{\kappa}=c^{\mrm{min}}_{p,q}$ and $h_{\kappa}=h_{p,q;2,1}$.

\section{Extension of SLE along the coset construction}
In this section, the main body of the present paper, we show how a generalization of SLE that has internal degrees of freedom is connected to WZW theory of $SU(2)$
from the viewpoint of coset construction.

\label{sect:SLE_WZW}
\subsection{Extended group $\mrm{Aut}_{+}\mcal{O}\ltimes G_{\mbb{C}}(\mfrak{m})$}
Firstly, we introduce a semi-direct product group denoted by $\mrm{Aut}_{+}\mcal{O}\ltimes G_{\mbb{C}}(\mfrak{m})$
and see that it acts on a completion of a representation of $\what{\mfrak{g}}$.
Let $G_{\mbb{C}}=SL(2,\mbb{C})$ be the complex Lie group corresponding to $\mfrak{g}$,
and let $G_{\mbb{C}}(\mcal{O})$ be the set of $\mcal{O}$-points in $G_{\mbb{C}}$, which admits the point wise product structure.
A relevant subgroup denoted by $G_{\mbb{C}}(\mfrak{m})$ consists of $\mfrak{m}$-points in $G_{\mbb{C}}$, where $\mfrak{m}\subset\mcal{O}$ is the maximal ideal,
then its Lie algebra is realized as $\mfrak{g}\otimes \mbb{C}[[\zeta^{-1}]]\zeta^{-1}$.
Since the group $\mrm{Aut}_{+}\mcal{O}$ acts on $G_{\mbb{C}}(\mfrak{m})$ by the transformation of the variable,
we can take their semi-direct product group $\mrm{Aut}_{+}\mcal{O}\ltimes G_{\mbb{C}}(\mfrak{m})$.

On a representation $L_{\mfrak{g},k}(j)$ of $\what{\mfrak{g}}$, the operator $L_{0}$ defined by the Sugawara construction 
is diagonalized so that $L_{\mfrak{g},k}(j)=\bigoplus_{n=0}^{\infty}L_{\mfrak{g},k}(j)_{h^{\mrm{Sug}}_{j}+n}$.
Then again we can consider the formal completion $\overline{L_{\mfrak{g},k}(j)}=\prod_{n=0}^{\infty}L_{\mfrak{g},k}(j)_{h^{\mrm{Sug}}_{j}+n}$.
We already know that this space admits an action of $\what{\mfrak{g}}$ and the Virasoro algebra
through the Sugawara construction;
but moreover, their commutation relation $[L_{m},X(n)]=-nX(m+n)$ ensures that
the semi-direct product Lie algebra $\mrm{Vir}\ltimes \what{\mfrak{g}}$ is acting on the same space.
It is also obvious that using this action, one can define a representation $\pi^{k}$ of
$\mrm{Aut}_{+}\mcal{O}\ltimes G_{\mbb{C}}(\mfrak{m})$ on $\overline{L_{\mfrak{g},k}(j)}$.

\subsection{Coupling of SLE with WZW theory}
As we have already seen, SLE($\kappa$) is coupled to CFT associated with the Virasoro algebra,
or more precisely, to the representation $L(c_{\kappa},h_{\kappa})$
via the property that $Q(f_{t})\ket{c_{\kappa},h_{\kappa}}$ is a local martingale,
where $f_{t}$ is the shifted SLE($\kappa$).
In this section, we consider the case when $\kappa=4p/(p+q)$ with coprime integers $p$ and $q$, and we regard the corresponding
representation $L(c^{\mrm{min}}_{p,p+q},h_{p,p+q;2,1})$ as one appearing in the coset construction,
and extend the notion of SLE with the guiding principle that
local martingales are computed from the representation theory.

In Theorem \ref{thm:GKO}, the desired representation $L(c^{\mrm{min}}_{p,p+q},h_{p,p+q;2.1})$
of the Virasoro algebra appears in the case when $k=-2+p/q$, $j=\epsilon=\frac{1}{2}$,
with the vacuum representation $L_{\mfrak{g},k+1}(0)$ of $\what{\mfrak{g}}$ of level $k+1$ as its partner.
Indeed the vector
\begin{equation}
	\ket{s}:=\Ket{\frac{1}{2}}_{k}\otimes F\Ket{\frac{1}{2}}_{1}-F\Ket{\frac{1}{2}}_{k}\otimes\Ket{\frac{1}{2}}_{1}
\end{equation}
is identified with $\ket{c^{\mrm{min}}_{p,p+q},h_{p,p+q;2,1}}\otimes \ket{0}_{k+1}\in L(c^{\mrm{min}}_{p,p+q},h_{p,p+q;2,1})\otimes L_{\mfrak{g},k+1}(0)$.
Here we put subscripts that specify the levels of representations in which the vectors lie.

Since conformal vectors in the current context are in relation
$\omega^{\mrm{Sug}}_{k}\otimes \ket{0}+\ket{0}\otimes\omega^{\mrm{Sug}}_{1}=\iota(\omega^{\mrm{Sug}}_{k+1})+\omega^{\mrm{Com}}$,
we have a relation among actions of the Virasoro algebra as follows:
\begin{equation}
	L^{\mfrak{g},k}_{n}+L^{\mfrak{g},1}_{n}=L^{\mrm{Vir}}_{n}+L^{\mfrak{g},k+1}_{n},
\end{equation}
where superscripts specify the spaces on which the operators act.
Since the action of $\mrm{Aut}_{+}\mcal{O}$ on a completed representation space
is defined by exponentiating the action of the Virasoro algebra, this relation implies
that the diagonal action of $\mrm{Aut}_{+}\mcal{O}$ can be expressed as
\begin{equation}
	\label{eq:diagonal_Aut}
	\pi^{k}(\rho)\otimes \pi^{1}(\rho)=Q^{\mrm{Vir}}(\rho)\otimes \pi^{k+1}(\rho)
\end{equation}
for $\rho\in\mrm{Aut}_{+}\mcal{O}$.
What is important here is that the diagonal action of the Virasoro algebra
cannot be written in terms of the diagonal action of the affine Lie algebra
because the action of the Virasoro algebra is defined by the Sugawara construction
that involves quadratic terms in the action of the affine Lie algebra.
On the other hand, the diagonal action of the affine Lie algebra is
by definition is written as the action of $\what{\mfrak{g}}$ of level $k+1$.
Since the action of $G_{\mbb{C}}(\mfrak{m})$ is given by exponentials of the action of the affine Lie algebra,
we can express the diagonal action of $G_{\mbb{C}}(\mfrak{m})$ as
\begin{equation}
	\label{eq:diagonal_G}
	\pi^{k}(\Theta)\otimes \pi^{1}(\Theta)=\mrm{Id}^{\mrm{Vir}}\otimes \pi^{k+1}(\Theta)
\end{equation}
for $\Theta\in G_{\mbb{C}}(\mfrak{m})$.
Here $\mrm{Id}^{\mrm{Vir}}$ is the identity operator on a representation space of the Virasoro algebra.
If we write $\scr{G}=\Theta\rho$ and combine Eq.(\ref{eq:diagonal_Aut}) and Eq.(\ref{eq:diagonal_G}),
we obtain an expression of the diagonal action of $\mrm{Aut}_{+}\mcal{O}\ltimes G_{\mbb{C}}(\mfrak{m})$:
\begin{equation}
	\pi^{k}(\scr{G})\otimes \pi^{1}(\scr{G})=Q^{\mrm{Vir}}(\rho)\otimes \pi^{k+1}(\scr{G}).
\end{equation}

We again consider the random process $f_{t}$ on $\mrm{Aut}_{+}\mcal{O}$
that is identified with the shifted SLE$(\kappa)$,
and a random process $\pi^{k}(f_{t})\otimes \pi^{1}(f_{t})\ket{s}$.
As we have already seen, if we take $\kappa=4p/(p+q)$,
then $Q^{\mrm{Vir}}(f_{t})\ket{c^{\mrm{min}}_{p.p+q},h_{p,p+q;2,1}}$ is a local martingale,
but since the partner $\pi^{k+1}(f_{t})\ket{0}_{k+1}$ gives a nontrivial contribution,
the total random process 
\begin{equation}
	\pi^{k}(f_{t})\otimes \pi^{1}(f_{t})\ket{s}=Q^{\mrm{Vir}}(f_{t})\ket{c^{\mrm{min}}_{p,p+q},h_{p,p+q;2,1}}\otimes \pi^{k+1}(f_{t})\ket{0}_{k+1}
\end{equation}
is not a local martingale.
Thus we need some modification to obtain a local martingale in this total space.

To motivate the SDE presented below, let us see how the increment
of the random process $\pi^{k+1}(f_{t})\ket{0}_{k+1}$ looks like.
Important points are that the vacuum vector $\ket{0}_{k+1}$ is translation invariant,
{\it i.e.}, $L^{\mfrak{g},k+1}_{-1}\ket{0}_{k+1}=0$,
and the action of the Virasoro algebra is given by the Sugawara construction.
Thus we see that
\begin{equation}
	\pi^{k+1}(f_{t})d\pi^{k+1}(f_{t})\ket{0}_{k+1}=-2L^{\mfrak{g},k+1}_{-2}\ket{0}_{k+1}dt=-\frac{1}{k+3}\sum_{a=1}^{3}X_{a}(-1)^{2}\ket{0}_{k+1}dt.	
\end{equation}
Then we wonder whether one can compensate this increment by considering random process
along the internal space.

To this aim, we consider a random process $\scr{G}_{t}=\Theta_{t}f_{t}$ on $\mrm{Aut}_{+}\mcal{O}\ltimes G_{\mbb{C}}(\mfrak{m})$,
where $f_{t}$ on $\mrm{Aut}_{+}\mcal{O}$ is identical to the one introduced above,
and $\Theta_{t}$ is on $G_{\mbb{C}}(\mfrak{m})$.
We impose the following SDE on $\Theta_{t}$:
\begin{equation}
	\Theta^{-1}_{t}d\Theta_{t}=\frac{\tau}{2}\sum_{a=1}^{3}(X_{a}\otimes f_{t}(\zeta)^{-1})^{2}dt+\sqrt{\tau}\sum_{a=1}^{3}X_{a}\otimes f_{t}(\zeta)^{-1} dW_{t}^{(a)}
\end{equation}
Here $W_{t}^{(a)}$ for $a=1,2,3$ are mutually independent standard Brownian motions
that are also independent of $B_{t}$ and start at the origin.
Then the random process $\scr{G}_{t}$ satisfies the SDE
\begin{align}
	\scr{G}_{t}^{-1}d\scr{G}_{t}=
	&\left(-2\ell_{-2}+\frac{\kappa}{2}\ell_{-1}^{2}+\frac{\tau}{2}\sum_{a=1}^{3}X_{a}(-1)^{2}\right)dt \\
	&+\sqrt{\kappa}\ell_{-1}dB_{t}+\sqrt{\tau}\sum_{a=1}^{3}X_{a}(-1)dW_{t}^{(a)}. \notag
\end{align}
By definition of the Sugawara construction, $\pi^{k+1}(\scr{G}_{t})\ket{0}_{k+1}$
is a local martingale if we take $\tau=\frac{2}{k+3}$.
Note that the martingale condition on $\pi^{k+1}(\scr{G}_{t})\ket{0}_{k+1}$
does not cause any restriction on the parameter $\kappa$
because the vacuum vector is translation invariant.
The translation invariance of the vacuum vector also ensures that the tensor product
$Q^{\mrm{Vir}}(f_{t})\ket{c^{\mrm{min}}_{p,p+q},h_{p,p+q;2,1}}\otimes \pi^{k+1}(\scr{G}_{t})\ket{0}_{k+1}$
is a local martingale as well.
Consequently, we have the following:

\begin{thm}
Let $k=-2+p/q$ be an admissible level, and
let $\kappa$ and $\tau$ be chosen as $\kappa=4\frac{k+2}{k+3}$ and $\tau=\frac{2}{k+3}$.
Then the random process $\pi^{k}(\scr{G}_{t})\otimes \pi^{1}(\scr{G}_{t})\ket{s}$ is a local martingale.
\end{thm}

Moreover, by applying maps ${\bf 1}\otimes \bra{1/2}$ and ${\bf 1}\otimes \bra{1/2}E$ from
$\overline{L_{\mfrak{g},k}(1/2)}\otimes\overline{L_{\mfrak{g},1}(1/2)}$ to $ \overline{L_{\mfrak{g},k}(1/2)}$
on the local martingale $\pi^{k}(\scr{G}_{t})\otimes \pi^{1}(\scr{G}_{t})\ket{s}$,
we also obtain the following result:

\begin{cor}
Let $k=-2+p/q$ be an admissible level.
With the same parameters $\kappa=4\frac{k+2}{k+3}$ and $\tau=\frac{2}{k+3}$,
$\pi^{k}(\scr{G}_{t})\ket{v_{\frac{1}{2}}}$ is a local martingale
for arbitrary choice $\ket{v_{\frac{1}{2}}}$ from the fundamental (spin-$\frac{1}{2}$)
representation $L(\frac{1}{2})=\mbb{C}\ket{1/2}\oplus F\ket{1/2}$.
\end{cor}

The construction of the martingale in Eq.(\ref{eq:martingale_WZW}) is almost in the same way as
in the case of the Virasoro algebra except for that
a primary field $\Psi(\ket{\lambda},z)$ for the $SU(2)$ WZW model transforms under the adjoint
action by $\pi^{k}(\Theta \rho)$ for $\rho\in \mrm{Aut}_{+}\mcal{O}$ and $\Theta\in G_{\mbb{C}}(\mfrak{m})$
so that
\begin{equation}
	\Psi(\ket{\lambda},z)= \pi^{k}(\Theta \rho)\Psi(\Theta^{-1}(z)\ket{\lambda},\rho(z))\pi^{k}(\Theta \rho)^{-1}(\pr{\rho}(z))^{h_{\lambda}},
\end{equation}
where $\Theta^{-1}(z)$ is a $\mbb{C}[[z^{-1}]]z^{-1}$-point in $SL(2,\mbb{C})$ obtained by
identifying the formal variable $\zeta$ in $\Theta^{-1}$ with the formal variable $z$.
If we write $\ket{\lambda(z)}_{t}$ for $\Theta_{t}^{-1}(z)\ket{\lambda}$,
we can verify that this random process satisfies the SDE in Eq.(\ref{eq:SDE_internal}).
Then the quantity in Eq.(\ref{eq:martingale_WZW}) is identified with
\begin{equation}
	\mcal{M}_{t}^{G}=
	\braket{0|\Psi(\ket{\lambda_{\infty}},\infty)\Psi(\ket{\lambda_{1}},z_{1})\cdots \Psi(\ket{\lambda_{n}},z_{n})\pi^{k}(\scr{G}_{t})|v_{\frac{1}{2}}},
\end{equation}
where $\ket{v_{\frac{1}{2}}}\in L(\frac{1}{2})$.
Then we have an alternative simpler proof of the following theorem presented
in the previous works \cite{BettelheimGruzbergLudwigWiegmann2005,AlekseevBytskoIzyurov2011}:
\begin{thm}
Let $k=-2+p/q$ be an admissible level.
For the choice of the parameters $\kappa=4\frac{k+2}{k+3}$ and $\tau=\frac{2}{k+3}$,
$\mcal{M}_{t}^{G}$ in Eq.(\ref{eq:martingale_WZW}) with $\lambda$ being the fundamental weight is a local martingale.
\end{thm}

\section{Conclusion}
\label{sect:conclusion}
In this paper, we investigated how SLE was coupled to WZW model for $SU(2)$
relying on the coset construction in \cite{GoddardKentOlive1986,KacWakimoto1988}.
In the classical theory of the SLE/CFT correspondence,
local martingales associated with SLE can be realized as a random process
on a completion of a representation of the Virasoro algebra
due to existence of a null vector.
The starting point of our present work was to regard this representation of the Virasoro algebra
as one obtained by the coset construction.
Importantly, the partner representation of $\what{\mfrak{g}}$ of level $k+1$ to the relevant representaion
of the Virasoro algebra was the vacuum representation.
Because of the translation invariance of the vacuum vector,
the martingale condition for the random process on the partner space
did not cause any restriction on the parameter $\kappa$,
but only did on the parameter $\tau$.
This can be interpreted that SLE and a random process in the internal space
are decoupled into the components of the tensor product space,
and explains how certain parameters are chosen in terms of the level of the WZW model.

From the viewpoint of the current work,
among generalizations of SLE corresponding to WZW theories,
the one for $SU(2)$ has a special importance
in the sense that the $\mcal{W}$-algebra corresponding to $\mfrak{sl}_{2}$
is the Virasoro algebra, which is connected to SLE.
For a simply laced finite dimensional simple Lie algebra $\mfrak{g}$,
it was proved in \cite{ArakawaCreutzigLinshaw2018} that
\begin{equation}
	\mcal{W}_{-h^{\vee}+\frac{k+h^{\vee}}{k+h^{\vee}+1}}(\mfrak{g})\simeq \mrm{Com}(L_{\mfrak{g},k+1},L_{\mfrak{g},k}\otimes L_{\mfrak{g},1})	,
\end{equation}
where $L_{\mfrak{g},k}$ is the affine vertex algebra for $\mfrak{g}$ of an admissible level $k$
defined in the same way as in the case of $\mfrak{sl}_{2}$,
$\mcal{W}_{k}(\mfrak{g})$ is the principal $\mcal{W}$-algebra for $\mfrak{g}$ of level $k$
defined by the quantum Drinfeld-Sokolov reduction,
and $h^{\vee}$ is the dual Coxeter number of $\mfrak{g}$.
Thus a possible future direction is to consider a generalization of SLE
that is connected to a $\mcal{W}$-algebra,
and in relation to this generalization,
discuss a generalization of SLE corresponding to a WZW theory.
Note that a $\mcal{W}$-algebra is not a Lie algebra but just a vertex algebra in general,
which might cause a difficulty in consideration of such a generalization
because a description using an infinite dimensional Lie group is no longer available.

We can also think of another method to couple SLE with a WZW model
via the coset construction proposed by Kac and Wakimoto \cite{KacWakimoto1990} and studied by Bouwknegt \cite{Bouwknegt1997}.
In this construction, one consider an embedding $\mfrak{g}_{jk}\to\mfrak{g}_{k}$ defined by $X(-1)\mapsto X(-j)$.
Then it has been shown that a representation $L_{\mfrak{g},k}(\Lambda)$ of $\what{\mfrak{g}}$
of weight $\Lambda$ and level $k$ is decomposed into representations of $\mrm{Vir}\oplus\what{\mfrak{g}}$.
This decomposition has been classified in the case when the central charge
for the Virasoro algebra is less than unity.
Possibly important examples include the case when $\mfrak{g}=E_{8}$, $k=1$ and $j=2$.
In this case, the representation $L_{E_{8},1}(0)$ has a component $L^{\mrm{Vir}}_{c^{\mrm{min}}_{3,4}}(h_{3,4;2,1})\otimes L_{E_{8},2}(0)$ in the decomposition,
where the representation of the Virasoro algebra is connected to SLE$(3)$ and the partner representation of
$\what{E_{8}}$ of level 2 is the vacuum representation.
Other relevant examples can be found in the work by Bouwknegt \cite{Bouwknegt1997}.
It will be interesting to formulate a generalization of SLE corresponding to a WZW theory
through this coset construction.

%
%
%
%
%
%
%
%




\addcontentsline{toc}{chapter}{Bibliography}
\bibliographystyle{alpha}
\bibliography{sle_cft}

\newcommand{\etalchar}[1]{$^{#1}$}
\begin{thebibliography}{CDCH{\etalchar{+}}14}

\bibitem[ABI11]{AlekseevBytskoIzyurov2011}
A.~Alekseev, A.~Bytsko, and K.~Izyurov.
\newblock On {SLE} martingales in boundary {WZW} models.
\newblock {\em Lett. Math. Phys.}, 97:243--261, 2011.

\bibitem[ACL18]{ArakawaCreutzigLinshaw2018}
T.~Arakawa, T.~Creutzig, and R.~Linshaw.
\newblock $w$-algebras as coset vertex algebras, 2018.
\newblock arXiv:1801.03822.

\bibitem[BB02]{BauerBernard2002}
M.~Bauer and D.~Bernard.
\newblock {SLE}$_{\kappa}$ growth processes and conformal field theories.
\newblock {\em Phys. Lett. B}, 543:135--138, 2002.

\bibitem[BB03a]{BauerBernard2003a}
M.~Bauer and D.~Bernard.
\newblock Conformal field theories of stochastic {Loewner} evolutions.
\newblock {\em Commun. Math. Phys.}, 239:493--521, 2003.

\bibitem[BB03b]{BauerBernard2003b}
M.~Bauer and D.~Bernard.
\newblock {SLE} martingales and the {Virasoro} algebra.
\newblock {\em Phys. Lett. B}, 557:309--316, 2003.

\bibitem[BBK05]{BauerBernardKytola2005}
M.~Bauer, D.~Bernard, and K.~Kyt{\"o}l{\"a}.
\newblock Multiple {Schramm-Loewner} evolutions and statistical mechanics
  martingales.
\newblock {\em J. Stat. Phys.}, 120:1125--1163, 2005.

\bibitem[BGLW05]{BettelheimGruzbergLudwigWiegmann2005}
E.~Bettelheim, I.~A. Gruzberg, A.~W.~W. Ludwig, and P.~Wiegmann.
\newblock Stochastic {Loewner} evolution for conformal field theories with
  {Lie} group symmetries.
\newblock {\em Phys. Rev. Lett.}, 95:251601, 2005.

\bibitem[Bou97]{Bouwknegt1997}
P.~Bouwknegt.
\newblock Coset construction for winding subalgebras and applications, 1997.
\newblock arXiv:q-alg/9610013.

\bibitem[BPZ84]{BelavinPolyakovZamolodchikov1984}
A.~A. Belavin, A.~M Polyakov, and A.~B. Zamolodchikov.
\newblock Infinite conformal symmetry in two-dimensional quantum field theory.
\newblock {\em Nucl. Phys. B}, 241:333--380, 1984.

\bibitem[Car86]{Cardy1986}
J.~L. Cardy.
\newblock Effect of boundary conditions on the operator center of
  two-dimensional conformally invariant theories.
\newblock {\em Nucl. Phys. B}, 275:200--218, 1986.

\bibitem[Car89]{Cardy1989}
J.~L. Cardy.
\newblock Boundary conditions, fusion rules and the {Verlinde} formula.
\newblock {\em Nucl. Phys. B}, 324:581--596, 1989.

\bibitem[Car92]{Cardy1992}
J.~L. Cardy.
\newblock Critical percolation in finite geometries.
\newblock {\em J. Phys. A: Math. Gen.}, 25:L201--L206, 1992.

\bibitem[CDCH{\etalchar{+}}14]{ChelkakDuminil-CopinHonglerKemppainenSmirnov2014}
D.~Chelkak, H.~Duminil-Copin, C.~Hongler, A.~Kemppainen, and S.~Smirnov.
\newblock Convergence of {Ising} interfaces to {Schramm's SLE} curves.
\newblock {\em Comptes Rendus Mathematique}, 352:157--161, 2014.

\bibitem[Dub15a]{Dubedat2015b}
J.~Dub{\'e}dat.
\newblock {SLE} and {Virasoro} representations: Fusion.
\newblock {\em Commun. Math. Phys.}, 336:761--809, 2015.

\bibitem[Dub15b]{Dubedat2015a}
J.~Dub{\'e}dat.
\newblock {SLE} and {Virasoro} representations: Localization.
\newblock {\em Commun. Math. Phys.}, 336:695--760, 2015.

\bibitem[FBZ04]{FrenkelBen-Zvi2004}
E.~Frenkel and D.~Ben-Zvi.
\newblock {\em Vertex Algebras and Algebraic Curves}, volume~88 of {\em
  Mathematical Surveys and Monographs}.
\newblock American Methematical Society, 2nd edition, 2004.

\bibitem[FK04]{FriedrichKalkkinen2004}
R.~Friedrich and J.~Kalkkinen.
\newblock On conformal field theory and stochastic {Loewner} evolution.
\newblock {\em Nucl. Phys. B}, 687:279--302, 2004.

\bibitem[Fri04]{Friedrich2004}
R.~Friedrich.
\newblock On connections of conformal field theory and stochastic {Loewner}
  evolution, 2004.
\newblock arXiv:math-ph/0410029.

\bibitem[FW03]{FriedrichWerner2003}
R.~Friedrich and W.~Werner.
\newblock Conformal restriction, highest-weight representations and {SLE}.
\newblock {\em Commun. Math. Phys.}, 243:105--122, 2003.

\bibitem[FZ92]{FrenkelZhu1992}
I.~B. Frenkel and Y.~Zhu.
\newblock Vertex operator algebras associated to representations of affine and
  {Virasoro} algebras.
\newblock {\em Duke Math. J.}, 66:123--168, 1992.

\bibitem[GKO86]{GoddardKentOlive1986}
P.~Goddard, A.~Kent, and P.~Olive.
\newblock Unitary representations of the virasoro and super-virasoro algebras.
\newblock {\em Commun. Math. Phys.}, 103:105--119, 1986.

\bibitem[IK11]{IoharaKoga2011}
K.~Iohara and Y.~Koga.
\newblock {\em Representation Theory of the Virasoro Algebra}.
\newblock Springer Monographs in Mathematics. Springer-Verlag London, 2011.

\bibitem[Kac98]{Kac1998}
V.~Kac.
\newblock {\em Vertex Algebras for Beginners}, volume~10 of {\em University
  Lecture Series}.
\newblock American Mathematical Society, Providence, RI, 2nd edition, 1998.

\bibitem[Kon03]{Kontsevich2003}
M.~Kontsevich.
\newblock {CFT, SLE} and phase boundaries, 2003.
\newblock Oberwolfach Arbeitstagung.

\bibitem[Kos17]{SK2017}
S.~Koshida.
\newblock {SLE}-type growth processes corresponding to {Wess-Zumino-Witten}
  theories, 2017.
\newblock arXiv:1710.03835.

\bibitem[Kos18a]{SK2018a}
S.~Koshida.
\newblock Local martingales associated with {SLE} with internal symmetry, 2018.
\newblock arXiv:1803.06808.

\bibitem[Kos18b]{SK2018c}
S.~Koshida.
\newblock Note on {Schramm-Loewner} evolution for superconformal algebras,
  2018.
\newblock arXiv:1805.02891.

\bibitem[Kos18c]{SK2018b}
S.~Koshida.
\newblock {Schramm-Loewner evolution} with {Lie} superalgebra symmetry.
\newblock {\em Int. J. Mod. Phys.}, 33:1850117, 2018.
\newblock arXiv:1803.09579.

\bibitem[KS07]{KontsevichSuhov2007}
M.~Kontsevich and Y.~Suhov.
\newblock On {Malliavin} measures, {SLE}, and {CFT}.
\newblock {\em Proceedings of the Steklov Institute of Mathematics},
  258:100--146, 2007.

\bibitem[KW88]{KacWakimoto1988}
V.~G. Kac and M.~Wakimoto.
\newblock Modular invariant representations of infinite dimensional lie
  algebras and superalgebras.
\newblock {\em Proc. Natl. Acad. Soc.}, 35:4956--4960, 1988.

\bibitem[KW89]{KacWakimoto1989}
V.~Kac and M.~Wakimoto.
\newblock Classification of modular invariant representations of affine
  algebras.
\newblock In {\em Infinite-dimensional {Lie} algebras and groups}, volume~7 of
  {\em Adv. Ser. Math. Phys.}, pages 138--177. World Sci. Publ., Teaneck, NJ,
  1989.

\bibitem[KW90]{KacWakimoto1990}
V.~G. Kac and M.~Wakimoto.
\newblock Branching functions for winding subalgebras and tensor products.
\newblock {\em Acta Appl. Math.}, 21:3, 1990.

\bibitem[Kyt07]{Kytola2007}
K.~Kyt{\"o}l{\"a}.
\newblock Vorasoro module structure of local martingales of {SLE} variants.
\newblock {\em Rev. Math. Phys.}, 5:455--509, 2007.

\bibitem[Law04]{Lawler2004}
G.~F. Lawler.
\newblock An introduction to the stochastic {Loewner} evolution.
\newblock In {\em Random Walks and Geometry}. De Gruyter, 2004.

\bibitem[LSW01a]{LawlerSchrammWerner2001a}
G.~F. Lawler, O.~Schramm, and W.~Werner.
\newblock Values of brownian intersection exponents, {I}: Half-plane exponents.
\newblock {\em Acta Math.}, 187:237--273, 2001.

\bibitem[LSW01b]{LawlerSchrammWerner2001b}
G.~F. Lawler, O.~Schramm, and W.~Werner.
\newblock Values of brownian intersection exponents, {II}: Plane exponents.
\newblock {\em Acta Math.}, 187:275--308, 2001.

\bibitem[LSW02a]{LawlerSchrammWerner2002b}
G.~F. Lawler, O.~Schramm, and W.~Werner.
\newblock Analyticity of intersection exponents for planar brownian motion.
\newblock {\em Acta Math.}, 189:179--201, 2002.

\bibitem[LSW02b]{LawlerSchrammWerner2002a}
G.~F. Lawler, O.~Schramm, and W.~Werner.
\newblock Values of brownian intersection exponents, {III}: Two-sided
  exponents.
\newblock {\em Ann. Inst. H. Poincare (B) Probability and Statistics},
  38:109--123, 2002.

\bibitem[NR05]{NagiRasmussen2005}
J.~Nagi and J.~Rasmussen.
\newblock On stochastic evolutions and superconformal field theory.
\newblock {\em Nucl. Phys. B}, 704:475--489, 2005.

\bibitem[Ras04]{Rasmussen2004a}
J.~Rasmussen.
\newblock Stochastic evolutions in superspace and superconformal field theory.
\newblock {\em Lett. Math. Phys.}, 68:41--52, 2004.

\bibitem[RS05]{RohdeSchramm2005}
S.~Rohde and O.~Schramm.
\newblock Basic properties of {SLE}.
\newblock {\em Ann. Math.}, 161:883--924, 2005.

\bibitem[Sak13]{Sakai2013}
K.~Sakai.
\newblock Multiple {Schramm-Loewner} evolutions for conformal field theories
  with {Lie} algebra symmetries.
\newblock {\em Nucl. Phys. B}, 867:429--447, 2013.

\bibitem[Sch00]{Schramm2000}
O.~Schramm.
\newblock Scaling limits of loop-erased random walks and uniform spanning
  trees.
\newblock {\em Israel J. Math.}, 118:221--288, 2000.

\bibitem[Smi01]{Smirnov2001}
S.~Smirnov.
\newblock Critical percolation in the plane: conformal invariance, {Cardy's}
  formula, scaling limits.
\newblock {\em C. R. Acad. Sci. Paris}, 333:239--244, 2001.

\bibitem[Wer03]{Werner2003}
W.~Werner.
\newblock Random planar curves and {Schramm-Loewner} evolutions, 2003.
\newblock arXiv:math/0303354.

\end{thebibliography}



\end{document}